\documentclass{SciPost}

\hypersetup{
    colorlinks,
    linkcolor={red!50!black},
    citecolor={blue!50!black},
    urlcolor={blue!80!black}
}

\usepackage[bitstream-charter]{mathdesign}
\DeclareSymbolFont{usualmathcal}{OMS}{cmsy}{m}{n}
\DeclareSymbolFontAlphabet{\mathcal}{usualmathcal}

\fancypagestyle{SPstyle}{
\fancyhf{}
\lhead{\colorbox{scipostblue}{\bf \color{white} ~SciPost Physics Proceedings }}
\rhead{{\bf \color{scipostdeepblue} ~Submission }}

\fancyfoot[C]{\textbf{\thepage}}
}

\begin{document}

\pagestyle{SPstyle}

\begin{center}{\Large \textbf{\color{scipostdeepblue}{
Operational Challenges in Efficient and Sustainable Operation of the Event Processing Nodes GPU Farm at the ALICE Experiment\\
}}}\end{center}

\begin{center}\textbf{
Federico Ronchetti \textsuperscript{1$\star$},
Giada Erba\textsuperscript{2,3$\dagger$}
}\end{center}

\begin{center}
{\bf 1} CERN, Geneva, Switzerland
\\
{\bf 2} Goethe Universit\"at - Frankfurt am Main, Germany
\\
{\bf 3} Frankfurt Institute for Advanced Studies, Frankfurt am Main, Germany
\\[\baselineskip]
$\star$ \href{mailto:federico.ronchetti@cern.ch}{\small federico.ronchetti@cern.ch}\,,\quad
$\dagger$ \href{mailto:giada.erba@cern.ch}{\small giada.erba@cern.ch}
\end{center}

\definecolor{palegray}{gray}{0.95}
\begin{center}
\colorbox{palegray}{
  \begin{tabular}{rr}
  \begin{minipage}{0.14\textwidth}
    \includegraphics[width=25mm]{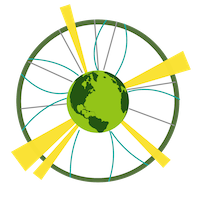}
  \end{minipage}
  &
  \begin{minipage}{0.58\textwidth}
    \vspace{5pt}
    \begin{center} \hspace{5pt}
    {\it Sustainable HEP workshop 2026 (SustHEP2026)} \\
    {\it Online, 08-10 July 2026
    }\\
    \doi{10.21468/SciPostPhysProc.?}
    \vspace{5pt}
    \end{center}

  \end{minipage}
\end{tabular}
}
\end{center}

\section*{\color{scipostdeepblue}{Abstract}}
\textbf{\boldmath{%
ALICE's Event Processing Nodes (EPN) farm is CERN's largest GPU computing
facility by capacity, comprising 350 GPU-based nodes and 2800 GPUs
delivering approximately a total of 48\,PFLOP/s. It has successfully enabled real-time
processing of heavy-ion collisions at 50\,kHz during LHC Run~3. The facility was designed for sustainability, extended
hardware lifetime, and reduced operational overhead. High availability is
supported by a robust technical architecture and well-defined operational
practices, ensuring reliable and continuous operation throughout data taking.
}}

\vspace{\baselineskip}

\noindent\textcolor{white!90!black}{%
\fbox{\parbox{0.975\linewidth}{%
\textcolor{white!40!black}{\begin{tabular}{lr}%
  \begin{minipage}{0.6\textwidth}%
    {\small Copyright attribution to authors. \newline
    This work is a submission to SciPost Phys. Proc. \newline
    License information to appear upon publication. \newline
    Publication information to appear upon publication.}
  \end{minipage} & \begin{minipage}{0.4\textwidth}
    {\small Received Date \newline Accepted Date \newline Published Date}%
  \end{minipage}
\end{tabular}}
}}
}




\section{Introduction}

During the second LHC long shutdown (LS\,2, 2019--2021), the ALICE experiment~\cite{ALICE:2022wpn} upgraded its detectors and computing infrastructure for Runs\,3 and 4.
A central component of the upgrade is the Online--Offline (O$^2$) framework~\cite{Buncic:2015ari}, which unifies data acquisition and offline reconstruction through continuous readout~\cite{alice_o2_continuous_readout}, and real-time processing.
To meet these requirements, ALICE deployed a dedicated CPU--GPU Event Processing Node (EPN) farm during LS\,2. GPUs provide high throughput for the highly parallel reconstruction tasks required at foreseen interaction rates, reducing the number of servers by roughly a factor of eight and lowering infrastructure and power costs.

\section{ALICE Computing for High-Rate Heavy-Ion Collisions}

The ALICE experiment has entered a new regime of high-rate heavy-ion data taking in Run\,3. With a 50\,ns bunch spacing, the instantaneous Pb--Pb luminosity reaches $\sim 6 \times 10^{27}\,\mathrm{cm}^{-2}\mathrm{s}^{-1}$, corresponding to a hadronic interaction rate of 50\,kHz. Coping with these rates required major detector and computing upgrades.



ALICE deployed a new computing model performs online calibration, event reconstruction, and data compression in real time. GPU acceleration is central to this approach since GPUs provide the highly parallel processing required for online tracking while reducing the number of physical nodes and the associated power consumption.

The operational core of this infrastructure is the EPN farm, comprising approximately 350 nodes equipped with 2800 GPUs. The farm processes the detector data stream during data taking and produces compressed data.

In Run\,3, ALICE collected an integrated Pb--Pb luminosity of $6.91\,\mathrm{nb}^{-1}$; the combined target for Runs\,3 and 4 is $13\,\mathrm{nb}^{-1}$.

\section{Online Reconstruction and Data Compression}

In the ALICE continuous-readout scheme, detector data are organized into Time Frames (TFs),
each representing a snapshot of raw data received from the readout system over a defined time interval~\cite{chep23}.
The data flow consists of a continuous stream of raw TFs transferred to the EPN farm for online processing.

The EPN farm performs tracking, calibration, reconstruction, and compression synchronously with data taking.
TPC tracking is by far the dominant online computing task, accounting for approximately 99\% of the GPU workload.
Calibration procedures are executed on a small number of dedicated CPU-only nodes, while TF compression exploits CPU vector instructions.

The raw TFs cannot be permanently stored. Instead, the EPN farm produces compressed Time Frames (CTFs), which are the only online data products retained for later processing.

\section{Hardware Accelerators for Sustainability}

GPU acceleration substantially reduces the required server count and energy
consumption.
The ALICE EPN farm relies extensively on this approach to meet the computational
requirements of online and offline event reconstruction. The system comprises
350 nodes equipped with 2,800 AMD Instinct GPUs. The resulting GPU-only capacity
provides approximately $43~\mathrm{PFLOP}$ of single-precision peak
performance, or $26~\mathrm{PFLOP}$ sustained.

In synchronous processing, where TPC tracking accounts for
approximately $99\%$ of the GPU workload, a single MI50 GPU replaces roughly
80 AMD Rome CPU cores.
In asynchronous processing, with around $80\%$ GPU
utilization, one MI50 replaces approximately 70 CPU cores. For comparison, a CPU-only
implementation of the EPN farm would therefore require more
than 2,000 64-core servers and would consume about twice as much energy.

280 MI50 nodes contain two AMD EPYC 7452 processors each, $512~\mathrm{GB}$ of
DDR4 memory, and eight $32~\mathrm{GB}$ AMD Instinct MI50 GPUs. 70 MI100
nodes are equipped with two AMD EPYC 7552 processors each, $1~\mathrm{TB}$ of DDR4
memory, and eight $32~\mathrm{GB}$ AMD Instinct MI100 GPUs. Nodes are
interconnected through $100~\mathrm{Gb/s}$ InfiniBand, complemented by
$1~\mathrm{Gb/s}$ Ethernet for management services. Four dedicated
calibration nodes provide additional CPU and memory resources for calibration
tasks.

\section{A Low-Impact Infrastructure}

The ALICE EPN farm is hosted in a modular, energy-efficient infrastructure
using adiabatic air cooling and on-site reverse osmosis water treatment.
Free cooling under favorable ambient conditions reduces energy consumption
and environmental impact. The facility achieves a PUE of 1.05--1.15, with
only 5--15\% of the total power used for cooling and other infrastructure
overheads.

The modular container-based design enables incremental expansion and
comprises IT and cooling containers, a dedicated power container with dual
$2.5~\mathrm{MW}$ transformers, a reverse-osmosis plant, and spare-equipment
facilities~\cite{epn-efficient-2025}. The infrastructure is operated and maintained by the EPN team,
with support from ALICE Technical Coordination, CERN IT, EN-CV, and EN-EL departments.
The EPN IT infrastructure life will be extended through Run\,4.
Beyond Run\,4, further upgrades (such as direct liquid cooling support)
could be required to support the next generation GPU-based farm.

\section{The Data Distribution System}
The EPN Data Distribution (DD)  \cite{neskovic2019data} system is responsible for collecting detector sub-Time Frames (sTFs), transporting them to the EPN farm, and assembling complete Time Frames (TFs). At the First-Level Processing (FLP) servers, incoming detector raw data are aggregated into sub-Time Frames (sTFs)
by the sTFBuilder which organizes the data blocks from the acquisition logic (CRU) over multiple LHC orbits ($88.93\mu\mathrm{s}$) into a single sTF.
Each sTF contains the data produced by one FLP (i.e. by a specific detector region) during a common time interval.
The sTFs are transferred from the FLPs to the EPN farm using Remote Direct Memory Access (RDMA) over InfiniBand (IB).

\section{Synchronous Processing}

During data taking, the ALICE detector pushes data with an aggregate input bandwidth of approximately $3.5\,\mathrm{TB/s}$. The FLP farm reduces this to about $0.7\,\mathrm{TB/s}$ (during Pb--Pb collisions) through local processing. The resulting data are transferred to the EPN farm, where calibration, reconstruction, and compression are performed synchronously.

The First-Level Processing (FLP) farm carries out local processing, however the most prominent task is the TPC zero suppression performed at the Common Readout Unit FPGA cards level.

The principal reconstruction workload is executed in the GPU barrel-processing chain. This chain performs TPC cluster finding, tracking, track merging, and distortion corrections, followed by the reconstruction and matching of a subset of tracks.

Online calibrations interact with the ALICE Experiment Control System (AliECS) \cite{Mrnjavac2024_AliECS}, the Detector Control System (DCS) \cite{alice_dcs_architecture_2013}, and the Conditions Database (CCDB) \cite{alice_ccdb_run3_2024}. This allows calibration and detector-condition information to be incorporated promptly into reconstruction.  The reconstructed data are finally passed to an entropy encoder, which produces CTFs \cite{ans-coding}. The EPN farm transforms the high-bandwidth detector stream into partially calibrated data. Tracking in the synchronous reconstruction is used for data compression as cluster rejection and entropy reduction. The output of synchronous processing (CTFs and calibration data) is transferred to the central
$150~\mathrm{PB}$ disk-storage system, which provides
$360~\mathrm{GB/s}$ throughput with approximately $25\%$ redundancy and makes it available to the ALICE Grid for asynchronous processing and analysis.

\section{Asynchronous Processing}

Following synchronous event processing, the EPN farm performs a full
calibration pass and asynchronous reprocessing to produce the final
reconstructed data.
Data reprocessing starts refining calibration data: more accurate data is produced by calibration passes using the latest constants available in CCDB.
The asynchronous stage then performs event reconstruction and produces Analysis Object Data (AOD).

CTFs are archived in the Tier-0 and Tier-1 storage systems while AOD are made available on the GRID for physics analysis.


\section{Run 3 EPN Performance}

During Run~3, the ALICE Event Processing Node (EPN) farm demonstrated
reliable synchronous processing under the demanding conditions of Pb--Pb
data taking. In 2024 and 2025, the instantaneous luminosity was
successfully leveled at 50~kHz, confirming that the online computing
infrastructure could sustain the expected interaction rates.

\begin{figure}[h]
    \centering
    \includegraphics[width=1.00\linewidth]{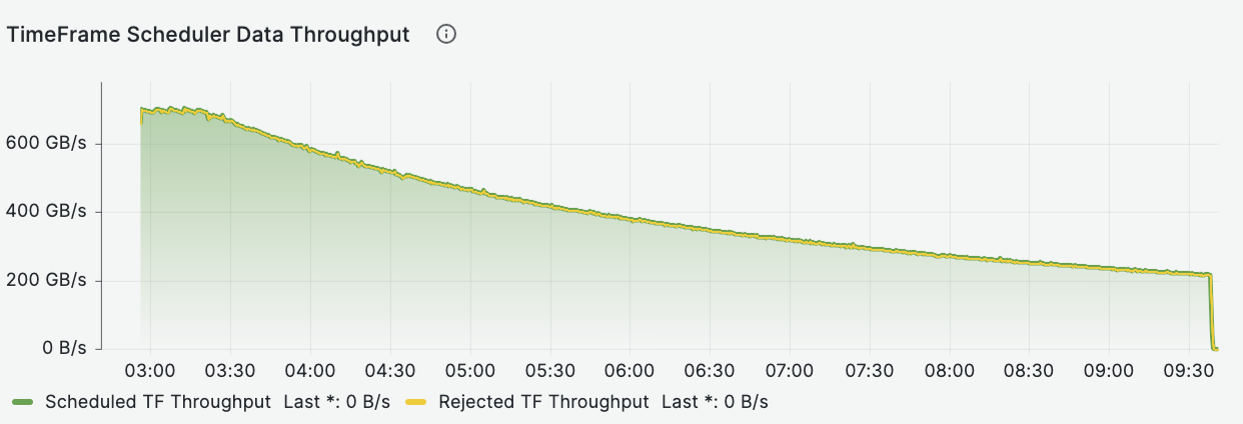}
  \caption{\small Dataflow in synchronous reconstruction during 2024 data taking. The initial flat part of the graph corresponds to the leveling at 50\,kHz of interaction rate which translates to $\sim 700$ GB/s of zero suppressed raw data into the ALICE EPN farm. The subsequent decrease follows the natural decay of the fill intensity. More detailed information on this plot is published in \cite{epn-efficient-2025}.}
    \label{fig:cr0}
\end{figure}



\section{Conclusions}

With the introduction of detector continuous readout, the successful operation of GPU-based synchronous reconstruction was essential to ALICE Run~3, enabling online calibration, reconstruction, and compression at the nominal 50~kHz Pb--Pb interaction rate and approximately 700~GB/s input to the reconstruction farm.
The use of an energy efficient air-to-air adiabatic cooling infrastructure and GPUs for online tracking was crucial to contain computing and environmental costs,
reducing the number of servers by a factor of eight and halving the energy consumption compared to a CPU-only implementation.

\section*{Acknowledgements}
The authors would like to thank the ALICE Technical Coordination, the CERN EN-CV, EN-EL, and IT departments for their support in the maintenance of the EPN IT infrastructure. We gratefully acknowledge the contributions of the ALICE O$^2$ PDP
(Physics Data Processing) team whose work and operational support were essential to the successful commissioning and operation of the EPN farm during Run\,3.

\paragraph{Author contributions}
F.R. contributed to the design, deployment, commissioning, and operation of the ALICE EPN GPU computing farm and IT infrastructure. He coordinated the integration and validation of the EPN farm within the O$^2$ framework, with particular responsibility for system performance, monitoring, optimization, and reliable operation during the initial phase of the Run\,3 data taking.
G.E. contributed to the deployment, configuration, and operation of the ALICE EPN computing farm, with primary responsibility for the system and network administration of the farm. She developed and maintained the automation tasks used for large-scale configuration and software deployment, and managed the core services supporting the farm. She contributed to the reliability and operational readiness of the EPN cluster throughout Run\,3 data taking.

\paragraph{Funding information}
The ALICE EPN project is funded by the German Federal Ministry of Education and Research (BMBF).

\bibliography{abstract-FR-GE-sust2026.bib}

\end{document}